\documentclass[twoside,singlecolumn,preprint,9pt]{article}
\usepackage{extsizes}
\usepackage[super,sort&compress,comma]{natbib} 
\usepackage[version=3]{mhchem}
\usepackage[left=1.5cm, right=1.5cm, top=1.785cm, bottom=2.0cm]{geometry}
\usepackage{balance}
\usepackage{times,mathptmx}
\usepackage{sectsty}
\usepackage{graphicx} 
\usepackage{lastpage}
\usepackage[format=plain,singlelinecheck=false,font={stretch=1.125,small,sf},labelfont=bf,labelsep=space]{caption}
\usepackage{float}
\usepackage{fancyhdr}
\usepackage{fnpos}
\usepackage[english]{babel}
\usepackage{array}
\usepackage{droidsans}
\usepackage{charter}
\usepackage[T1]{fontenc}
\usepackage[usenames,dvipsnames]{xcolor}
\usepackage{setspace}
\usepackage[compact]{titlesec}

\usepackage{soul}  
\usepackage{array} 
\usepackage[version=3]{mhchem} 
\usepackage{subfigure}
\usepackage{graphicx}
\usepackage{hyperref}
\newcommand{\slab}{Cu(111)+vacancy}
\newcommand{\water}{\ce{H2O}}

\newcommand{\ea}{E$_\text{a}$}

\newcommand{\degree}{$^{\circ}$}
\makeatletter
\newcommand*{\citenumns}[2][]{%
 \begingroup
 \let\NAT@mbox=\mbox
 \let\@cite\NAT@citenum
 \let\NAT@space\NAT@spacechar
 \let\NAT@super@kern\relax
 \renewcommand\NAT@open{}%
 \renewcommand\NAT@close{}%
 \cite[#1]{#2}%
 \endgroup
}
\makeatother
\usepackage{epstopdf}

\definecolor{cream}{RGB}{222,217,201}

\begin{document}

\pagestyle{fancy}
\thispagestyle{plain}

\makeFNbottom
\makeatletter
\renewcommand\LARGE{\@setfontsize\LARGE{15pt}{17}}
\renewcommand\Large{\@setfontsize\Large{12pt}{14}}
\renewcommand\large{\@setfontsize\large{10pt}{12}}
\renewcommand\footnotesize{\@setfontsize\footnotesize{7pt}{10}}
\makeatother

\renewcommand{\thefootnote}{\fnsymbol{footnote}}
\renewcommand\footnoterule{\vspace*{1pt}%
\color{cream}\hrule width 3.5in height 0.4pt \color{black}\vspace*{5pt}} 
\setcounter{secnumdepth}{5}

\makeatletter 
\renewcommand\@biblabel[1]{#1}            
\renewcommand\@makefntext[1]%
{\noindent\makebox[0pt][r]{\@thefnmark\,}#1}
\makeatother 
\renewcommand{\figurename}{\small{Fig.}~}
\sectionfont{\sffamily\Large}
\subsectionfont{\normalsize}
\subsubsectionfont{\bf}
\setstretch{1.125} 
\setlength{\skip\footins}{0.8cm}
\setlength{\footnotesep}{0.25cm}
\setlength{\jot}{10pt}
\titlespacing*{\section}{0pt}{4pt}{4pt}
\titlespacing*{\subsection}{0pt}{15pt}{1pt}
\fancyfoot{}
\fancyfoot[RO]{\footnotesize{\sffamily{1--\pageref{LastPage} ~\textbar  \hspace{2pt}\thepage}}}
\fancyfoot[LE]{\footnotesize{\sffamily{\thepage~\textbar\hspace{3.45cm} 1--\pageref{LastPage}}}}
\fancyhead{}
\renewcommand{\headrulewidth}{0pt} 
\renewcommand{\footrulewidth}{0pt}
\setlength{\arrayrulewidth}{1pt}
\setlength{\columnsep}{6.5mm}
\setlength\bibsep{1pt}

\makeatletter 
\newlength{\figrulesep} 
\setlength{\figrulesep}{0.5\textfloatsep} 

\newcommand{\topfigrule}{\vspace*{-1pt}%
\noindent{\color{cream}\rule[-\figrulesep]{\columnwidth}{1.5pt}} }

\newcommand{\botfigrule}{\vspace*{-2pt}%
\noindent{\color{cream}\rule[\figrulesep]{\columnwidth}{1.5pt}} }

\newcommand{\dblfigrule}{\vspace*{-1pt}%
\noindent{\color{cream}\rule[-\figrulesep]{\textwidth}{1.5pt}} }

\makeatother
\twocolumn[
  \begin{@twocolumnfalse}
\sffamily
{\centering
\begin{tabular}{m{0.5cm} p{16.5cm} m{1.0cm} }

& \noindent\LARGE{\textbf{{\it Ab initio} Investigation of Effect of Vacancy on 
Dissociation of Water Molecule on Cu(111) Surface}} & \\
\vspace{0.3cm} & \vspace{0.3cm}& \vspace{0.3cm} \\
 & \noindent\large{Vaibhav Kaware, \textit{$^{a,}$}\textit{$^{b}$}$^{\dag \ast}$
 and Kavita Joshi \textit{$^{a}$}$^{\ddag \ast}$} & \\
  & \noindent\normalsize{
  Water dissociation is a rate limiting step in many industrially important
  chemical reactions.  In this investigation, climbing image nudged elastic band
  (CINEB) method, within the framework of density functional theory, is used to
  report the activation energies (\ea) of water dissociation on Cu(111) surface with a
  vacancy. Introduction of vacancy results in a reduced coordination of the
  dissociated products, which facilitates their availability for reactions that
  involve water dissociation as an intermediate step. Activation energy for
  dissociation of water reduces by nearly 0.2 eV on Cu(111) surface with
  vacancy, in comparison with that of pristine Cu(111) surface.  We also find
  that surface modification of the Cu upper surface is one of the possible
  pathways to dissociate water when the vacancy is introduced.  Activation
  energy, and the minimum energy path (MEP) leading to the transition state
  remain same for various product configurations.  CINEB corresponding to
  hydrogen gas evolution is also performed which shows that it is a two step
  process involving water dissociation.  We conclude that the introduction of
  vacancy facilitates the water dissociation reaction, by reducing the
  activation energy by about 20\%.
} & \\
\end{tabular}
}
 \end{@twocolumnfalse} \vspace{0.6cm}
]
\renewcommand*\rmdefault{bch}\normalfont\upshape
\rmfamily
\section*{}
\vspace{-1cm}

\footnotetext{\textit{$^{a}$Physical and Materials Chemistry Division, CSIR-National Chemical Laboratory, Pune, India - 411008}}
\footnotetext{\textit{$^{b}$Department of Physics, Savitribai Phule Pune University, Ganeshkhind, India - 411007}}
\footnotetext{\textit{$^{\ddag}$k.joshi@ncl.res.in,} \textit{$^{\dag}$v.kaware@ncl.res.in, vaibhav.kaware@gmail.com}}

\section{Introduction}
Water is of immense importance in many
physio-chemical, biological, and chemical
reactions.\cite{waterincatalysisbook,wgskineticsreview,wgsmicrokinetic,methanolfundamentals,methanolsteamreformationreview,h2obioball,bagchi2013waterbook,h2obioproc}
It interacts with almost all solid surfaces that it comes in contact with, 
in one way or the other. It is ubiquitously present in naturally occurring
phenomena and is important in industrial chemical reactions like 
methanol synthesis, steam reformation, and the water gas shift reaction (WGSR), to name a
few.
\cite{waterincatalysisbook,wgskineticsreview,wgsmicrokinetic,methanolfundamentals,methanolsteamreformationreview}
It is also a strong candidate for production of hydrogen gas for the future technology of hydrogen
fuel cells.
Splitting water into hydrogen gas in contact with the electrode surfaces, 
holds the key to building more efficient and effective fuel cells.
\cite{natureartificialphotosynthesis,naturealtenergy,solarwatersplit,inorganicwatersplit,naturephotocells}
Water dissociation reaction serves as a stereotypical example for model
building of other complex reactions as well.\cite{watermetalifacenature} 
However, the simplicity of interaction of water with surfaces is only deceptive, since
it is this interaction that leads to formation of water structures on metal
surfaces that are far from being similar.\cite{watermetalifacenature} 
Since water is central to so many important chemical reactions, it is worth 
investigating
its behaviour on various surfaces in search of better catalysts for its reactions.
Interaction of
water with Cu surface is of special importance for the 
WGSR. WGSR is a process in which the poisonous CO gas reacts with water, and
gets converted into \ce{CO2} and hydrogen gas. The reaction is typically used
for production of hydrogen gas, and further into ammonia in industries
like fertilizers and petroleum.
Low temperature WGSR uses a mixture of oxides of Cu, Zn, Cr, Al, and others as
catalyst, in varying proportions. 
However, active component of the catalyst at low temperature, is
copper metal crystallite.\cite{wgskineticsreview} 
In spite of being well studied, there is no unanimously agreed mechanism for
the working of WGSR. Interestingly, dissociation of water remains a crucial step
in the WGSR, irrespective of the mechanism that drives it. 
Water dissociation becomes all the more important in WGSR and other reactions
involving water dissociation, since it is the rate limiting
step in all of them, along with other intermediate processes.\cite{gokhalecu111wgs,wgskineticsreview}
Thus, investigating water dissociation
in varied environments and on different surfaces, becomes an essential task.

Dissociation of water has been studied on various pristine surfaces,
computationally, as well as 
experimentally.\cite{watersurfthiel,watersurfrevisited,watermetalifacenature,wateradswettingsurfrevue}
It dissociates on different pristine surfaces with varying degree of ease,
wherein the activation energy for its dissociation varies from 0.15 eV on
Si(110) to  1.36 eV on Cu(111).\cite{h2oonsi001,gokhalecu111wgs}
It is observed that for almost all of these metal surfaces, 
water stabilizes parallel to the plane of surface atoms.
The reason being that the highest occupied molecular orbital,
$1b^1$, of water molecule is
perpendicular to its plane. Orbital $1b^1$ interacts with the surface atoms,
which orients it parallel to the slab surface.\cite{wateradswettingsurfrevue}
Cu(111) is special as far as its interaction with water is concerned.
Water is known to bind weakly to Cu(111).\cite{wateradswettingsurfrevue} It provides a good surface for
dissociation reaction, since it lies at the delicate balance between wetting 
and de-wetting behavior with water.
This means that the binding of de-wetting water on Cu(111) can be made
either stronger, or can be further weakened by modifying the surface appropriately. 
As far as catalytic surfaces are concerned, 
it is also known that introduction of structural defects like kinks and steps 
activates the metal surface chemically for reactions like 
\ce{NO} dissociation, \ce{O2} dissociation, and \ce{CO} oxidation, along with water
dissociation.\cite{nooxidationsteps,cooxidationaunp,langmuir,o2dissociation}
Dissociation of water has been carried out previously on pristine
Cu(111) surfaces, along with other pristine 
surfaces.\cite{waterflat,h2ocu111phatak,h2ocu111disspnas}
Faj{\'\i}n et. al. found that the presence of steps on Cu surfaces reduces the
activation energy of water dissociation reaction.\cite{wateroncustepsfajin}
They concluded that the low
coordination along the steps is responsible for this behaviour.
However, apart from steps and kinks, it is also possible to reduce the coordination 
on pristine surfaces by the creation of a vacancy defect.

In this work, we study water dissociation on Cu(111) surface in presence of a
single atom vacancy. Vacancy on the surface of pristine Cu(111) surface gives rise to non-uniformity,
unlike on pristine Cu(111),
that provides the reactants with a variety in coordination to choose from. 
We have studied the changes in adsorption
energies and geometries of the reactant and the products, the minimum energy path, and
the activation energies in these cases.
It is the aim of this study to verify the changes, if any, in activation energy
of water dissociation on copper (111) surface, with vacancy. We also study the
site dependence of reactant, and product molecules and the activation energy.
The study is expected to emulate not only the defective copper surface as
catalyst, but also copper nanoparticles that inevidently posses such defects,
which may be used to catalyse the reaction.
\section{\label{sec:compute}Computational Details}
Kohn-Sham formulation of density function theory (DFT) is used to calculate
total energies and to optimize various structures. Projector Augmented Wave pseudopotential
\cite{paw1,paw2} is used, with Perdew Burke Ernzerhof 
(PBE) \cite{pbe1} approximation for the exchange - correlation and generalized gradient
approximation,\cite{pbe2}  as implemented in the
plane wave code, Quantum Espresso (QE).\cite{quantumespresso}
Van der Waals type interaction is incorporated in the calculation of energies.\cite{vdw1,vdw2,vdw3}
Energy cutof\mbox{}f for plane-waves is kept at 47 Ry, and 221 Ry for charge
density. 
Solid Bulk Cu was relaxed in a variable cell to obtain lattice parameter of 3.62
\AA, which is in agreement with the reported experimental value of
3.62 \AA.{\cite{cualatwy}}
A slab of 6 layers of atoms along the direction (111) and of 2x2 surface coverage, was then
cut, and a single Cu atom vacancy was created. 
This slab was then relaxed, with all atoms free to move.
Convergence of total energy with respect to varying cutoffs, k-points and vacuum
along z-direction was confirmed. A monkhorst-pack grid of 4x4x1 k-points provided
for a precision of better than 0.02 eV in the energy differences.
A vacuum equivalent of 8 layers of unitcell, along (111) direction,
(16.864 \AA) was found to be sufficient to avoid image interaction in the
z-direction. Forces were converged below 0.001 eV/\AA~during all optimizations,
and below 0.1 eV/\AA~for all CINEB calculations. Nine intermediate images were used for all the
CINEB calculations initially.
Although CINEB calculations are known to be highly computation intensive, 
multiple such runs were performed with increased number of intermediate images,
in order to span multiple pathways of the reaction in as much detail as
possible, and to validate the initially found MEPs.
The dissociation energy is defined as
$E_{dissociation}$ = $E_{IS} - E_{FS}$, while the adsorption energy is
calculated as
$E_{adsorption} = E_{suf+adsorbant} - (E_{surf}+E_{adsorbant})$.
Here, $E_{IS}$ is the total energy of the reactant, slab+\water, $E_{FS}$ is the
total energy of the product, viz., slab + `dissociated water', 
$E_{slab+adsorbant}$ is the total energy of slab with the adsorbed water molecule,
$E_{slab}$ is the total energy of the standalone slab, and 
$E_{adsorbant}$ is the energy of adsorbed molecule in its free/gaseous state.
\begin{figure}
 \centering
 \includegraphics[scale=0.23] {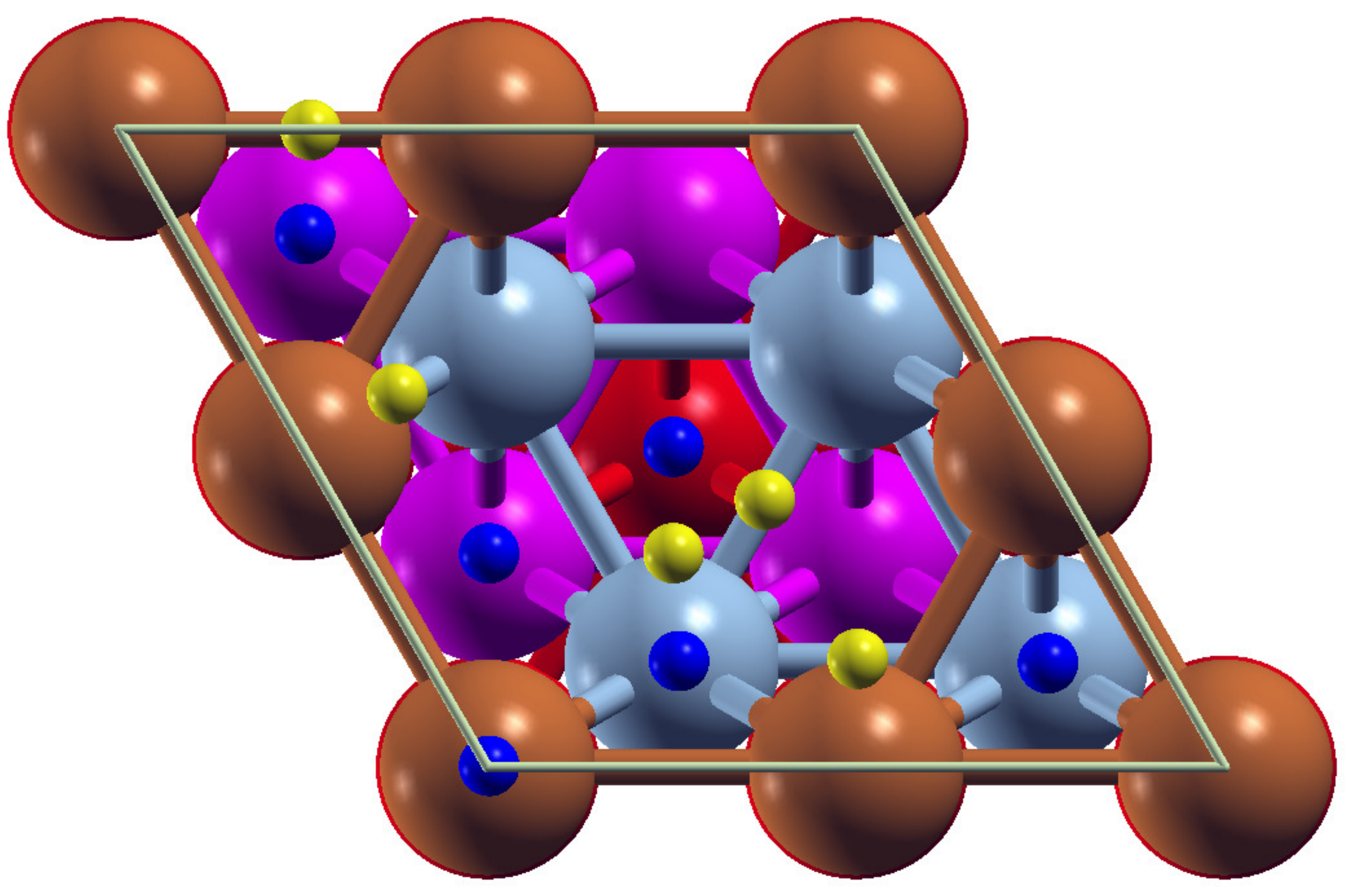}
 \label{fig:nonequivpos}
 \caption{(Color online) Non equivalent positions in a single cell. Vertex positions are
 colored blue, while bridge positions are colored yellow. Cu atoms in different
 planes are colored different in order to aid the eye. Brown colored spheres
 represent topmost layer `A' of Cu atoms, while grey spheres signify the Cu
 atoms in layer `B'. Magenta colored spheres are Cu atoms in layer `C', and 
 red spheres in layer D. The ABCABC stacking of Cu atom planes in pristine Cu 
 along (111) direction, is introduced with stacking plane D which is same as stacking
 plane A, but with a vacancy. Hence, stacking order is DBCABC for the 6 layers
 considered in these calculations.
}
 \label{fig:nonequivpos}
\end{figure}
\section{Results and Discussion}
As discussed in the introduction, water dissociation is a rate limiting step in 
many industrially important chemical reactions, like water gas shift reaction.
It is the aim of this work to study the effect of vacancy on the activation
energy of water dissociation on Cu(111). 
CINEB calculations are used to calculate the activation energies, along with the
MEPs.\cite{neb1,neb2,cineb} 
Nudged elastic band method is a chain of states method, used to find the 
MEP between two local minima on the potential energy surface.
Highest energy configuration along the MEP is the transition
state (TS), which is used to calculate the reaction rate, within the
harmonic transition state theory (hTST) approximation.
Climbing image method is used to promote the transition state, found by NEB,
to the saddle point.\cite{cineb}
Using CINEB calculations to find the MEP and the transition state requires that the 
two local minima, the initial (IS) and the final (FS) state, be known. 
In our case, the IS comprises of optimized slab+\water, while the FS is
`dissociated water'+slab. For the surface simulated here, there are more than
one available choices to place \water~molecule, and the product moieties. Hence,
calculations in this work were initiated with multiple optimizations for the IS as 
well as FS.
Eleven different non-equivalent positions were identified on the 2x2x6 slab,
consisting of six vertex and 5 bridge positions.
These positions are shown in Fig. \ref{fig:nonequivpos}.
Vertex positions are positions
in which the molecule was kept above the specific Cu atom site, and bridge
position implied placing the desired molecule between two Cu atomic sites.
Optimizations for initial (IS)  and the
final (FS) state were carried out on the fully relaxed Cu(111)+vacancy slab. The
optimization of Cu(111)+vacany slab resulted in its expansion along (111)
direction by about 1\%.
The vacancy site did not deform, expand or contract.
All atoms were kept free to move during all optimizations on the slab, and also
during the CINEB calculations.
\begin{figure}
\centering
 \subfigure[]{\includegraphics[scale=0.13] {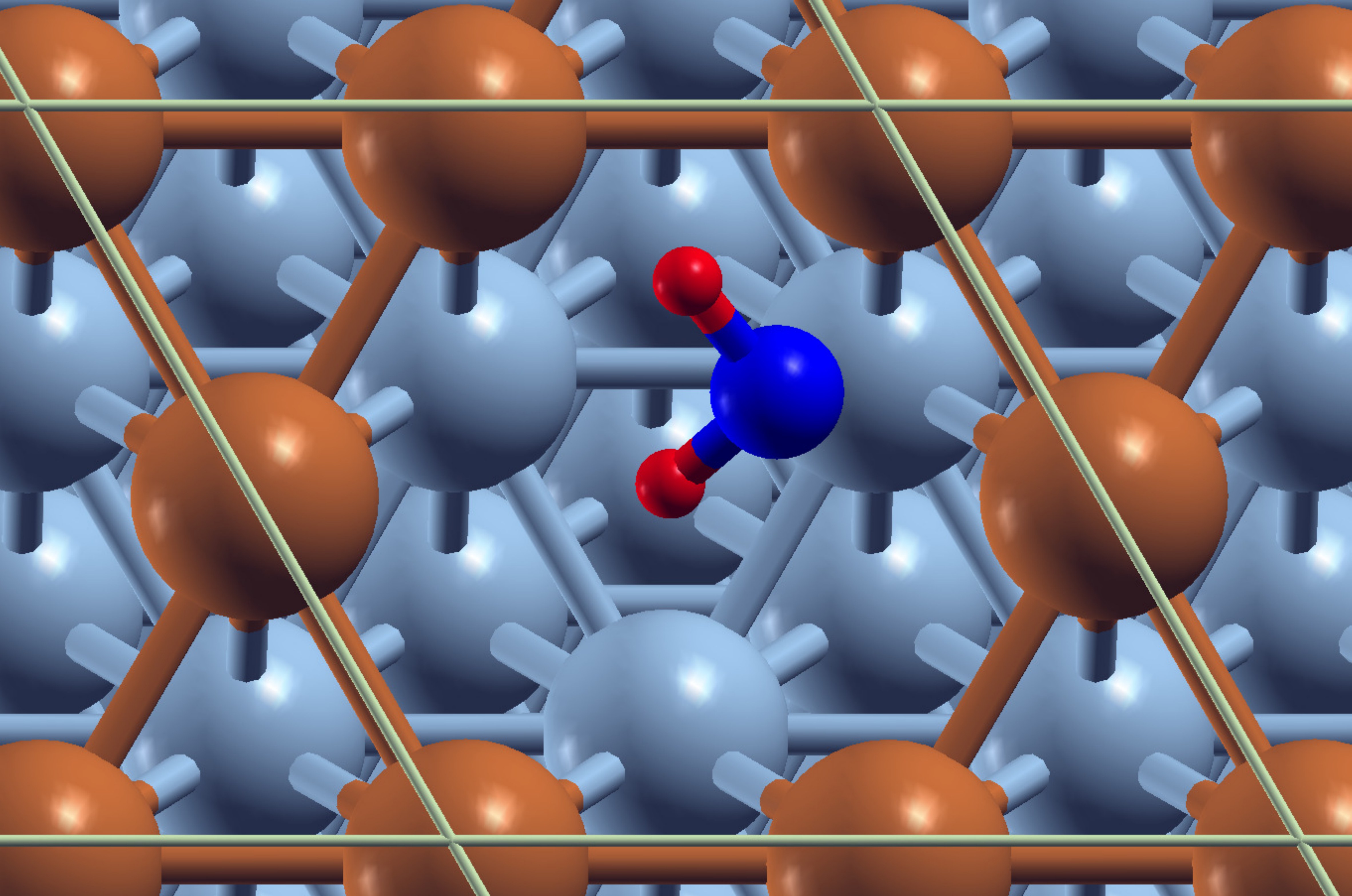}
 \label{fig:h2ovrtxafinal}}
 \subfigure[]{\includegraphics[scale=0.18] {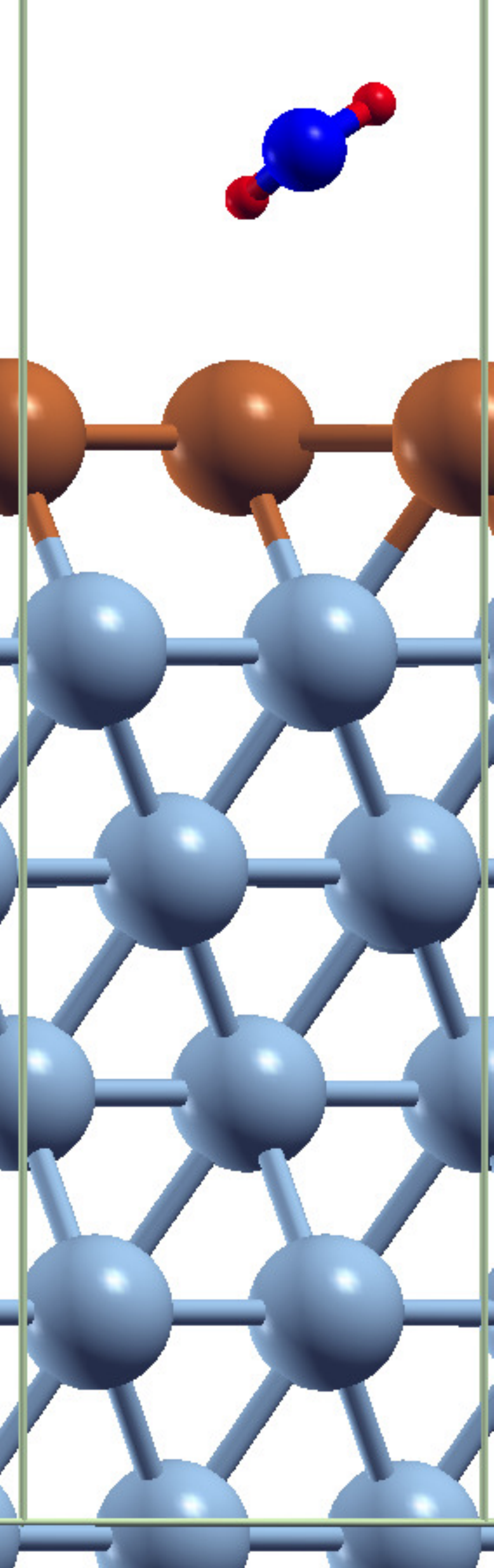}
 \label{fig:b}}
 \caption{(Color online) The optimized position of water molecule on the slab (a) along (111)
 z-direction, (b) lateral view. Topmost layer of copper atoms is signified by
 brown spheres, whereas the Cu atoms in layers below, are all colored as gray
 spheres for the visual aid. Red colored spheres represent the H atoms, while
 blue colored spheres stand for O atom.}
 \label{fig:is}
\end{figure}
\subsection{H$_2$O on \slab}
\begin{table}[!h]
   \begin{tabular}{
       m{2.37cm}| m{0.32cm}  m{0.35cm}   m{0.35cm}  m{0.89cm} m{0.55cm}
       m{0.32cm}  m{0.32cm}  } \hline
     Surface & E$_{ads}$ (eV) & $h$ (\AA) & d$_{HO}$ (\AA) &
     $\angle$\hspace{-0.00cm}HOH (\degree) &
     tilt ~ (\degree) & h$_{ads}$ (\AA)  & O$_{xy}$ (\AA) \\
    \hline
   pristine Cu(111) & 0.23 & 2.92 & 0.97 & 104 & 5.0    & 0.09 & 0.39 \\
    Cu(111)+vacancy & 0.26 & 2.83 & 0.97 & 104 & 37.66 &   -  &  -   \\
    \hline
   \end{tabular}
   \caption{Table of various distances and angles, with water~adsorbed on respective slab.
   tilt: Tilting angle of plane of water molecule with respect to the Cu[111] plane.
   $h$:Distance between \water~and
   slab. Value of $E_{ads}$ compares well with adsorption energy
   values of 0.21 eV from Ref.[\citenumns{h2ocu111phatak}],  0.18 eV from 
   Ref.[\citenumns{gokhalecu111wgs}], and 0.24 eV from
   Ref.[\citenumns{wateradswettingsurfrevue}].
   All other parameters compare well with those from Ref.[\citenumns{h2ocu111phatak}].}
   \label{table:wateradsorbtable}
\end{table}
For the initial state of reaction, water
molecule was kept at pre-designated distinct vertex sites (shown in
Fig. \ref{fig:nonequivpos}), and the bridge
sites, and their optimizations were carried out. 
It is reported that pristine Cu(111) aligns the \water~molecule parallel to (111)
direction.\cite{waterflat,gokhalecu111wgs,h2ocu111disspnas}
Our calculations agree to the same, where water molecule aligns parallel to the
(111) planes slightly off the atop position of Cu atom on pristine Cu(111)
surface. 
Upon introduction of the vacancy, we found that irrespective
of the initial position of water molecule, it always ended up lining at the
vacancy site, with a slight inward tilt. 
With the vacancy in place, \water~ cannot stay entirely
parallel to Cu(111) slab, and tilts downwards towards the slab. (Fig. \ref{fig:is}).
This is the degenerate global minimum energy configuration for water molecule to
get adsorbed on the Cu(111)+vacancy slab. 
Slightly different angles and/or
positions of water molecule at the vacancy result into degenerate energy
configurations. 
\water~stays adsorbed on the surface at $\sim$2.8 \AA~with adsorption energy of
0.26 eV. 
Comparison of the relevant parameters for water adsorbed on pristine and
Cu(111)+vacancy, is produced in Table \ref{table:wateradsorbtable}.
Cases where the water molecule was
kept at non-vacancy (pristine) site, the water molecule optimized into plane parallel to
Cu(111) slab, as it would in case of the entire pristine Cu(111) slab.
Adsorption energy of water molecule showed no change, 
whether the molecule ended up at the vacancy or at the pristine position.
The reason behind this behaviour is that
since water molecule is loosely adsorbed (via van der Waals interaction) on the Cu
surface, changes in Cu surface may not strongly reflect in water
adsorption.\cite{wateradswettingsurfrevue}
However, modifying the surface may be expected to modify the behaviour of 
dissociated moieties, since they form chemical bonds with Cu atoms of the
surface.
Thus, in summary, water molecule on Cu(111)+vacancy slab optimizes in a degenerate
position, where it tilts near the vacancy site. If placed upon high coordinated site,
it prefers to stay planar, parallel to (111) direction, but its adsorption energy,
does not change in either cases. 
\begin{figure}[!h]
  \centering
  \includegraphics[scale=0.70]{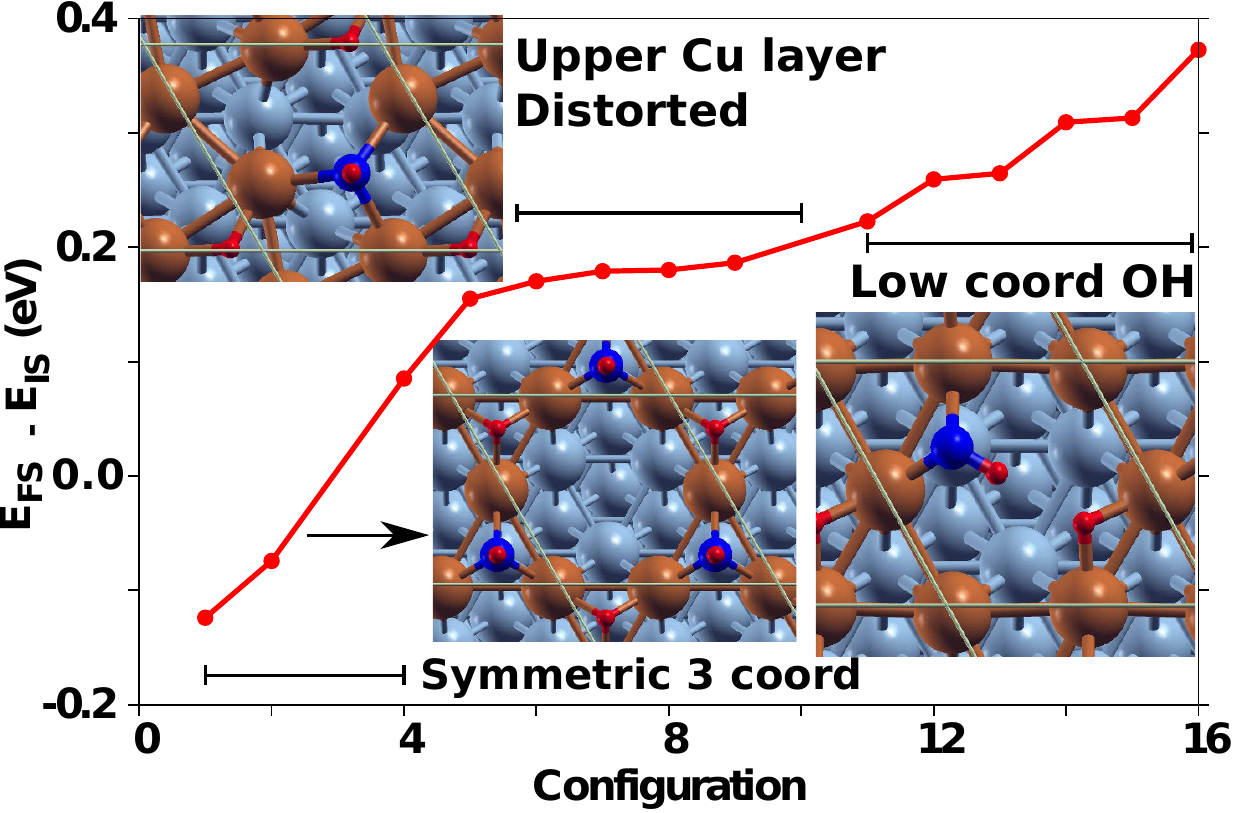}
  \caption{(Color online) Various H+OH optimized configurations on Cu(111)+vacancy slab
  placed on the graph of their energies with respect to the IS energy.
  Coloring convention of spheres is same as in Fig. \ref{fig:is}.
  }
  \label{fig:vcnohhenergies}
\end{figure}
\subsection{H+O+H on \slab}
\label{subsec:classifyFS}
\begin{figure*}
 \subfigure[]{
  \includegraphics[scale=0.255]{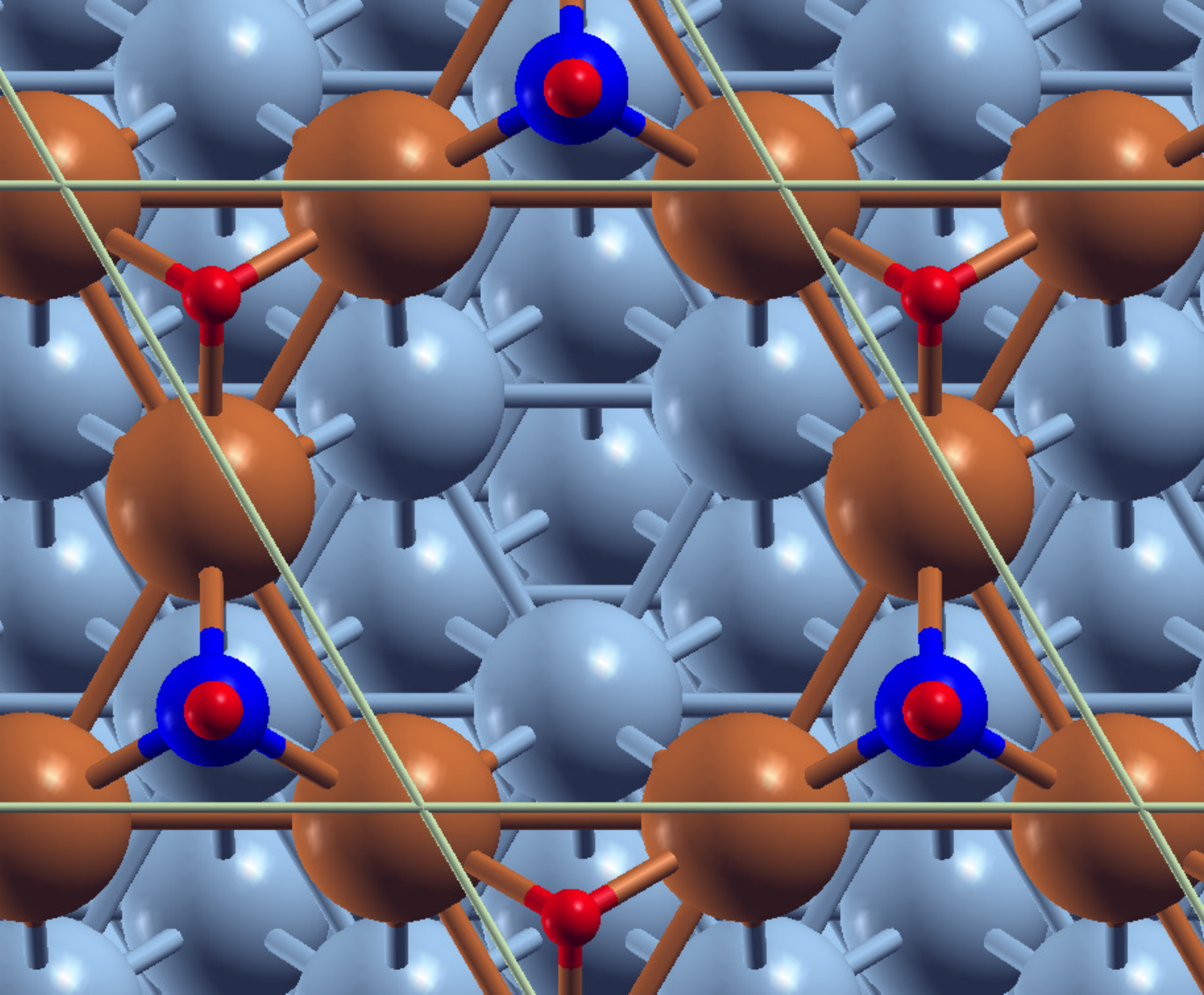}
  \label{fig:ohhmostcommon}
  }
 \subfigure[]{
  \includegraphics[scale=0.213]{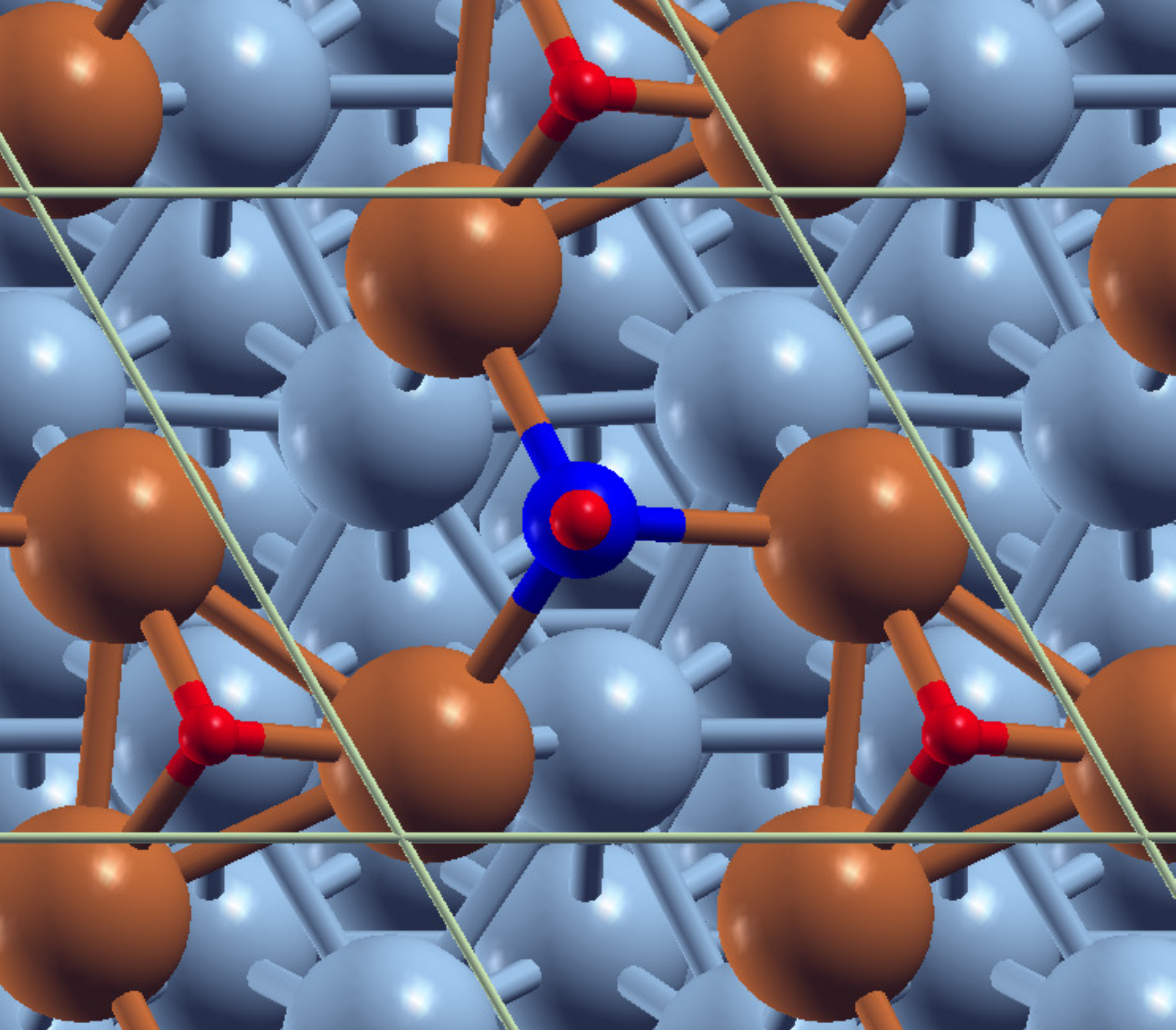}
  \label{fig:ohhdistort}
  }
 \subfigure[]{
  \includegraphics[scale=0.214]{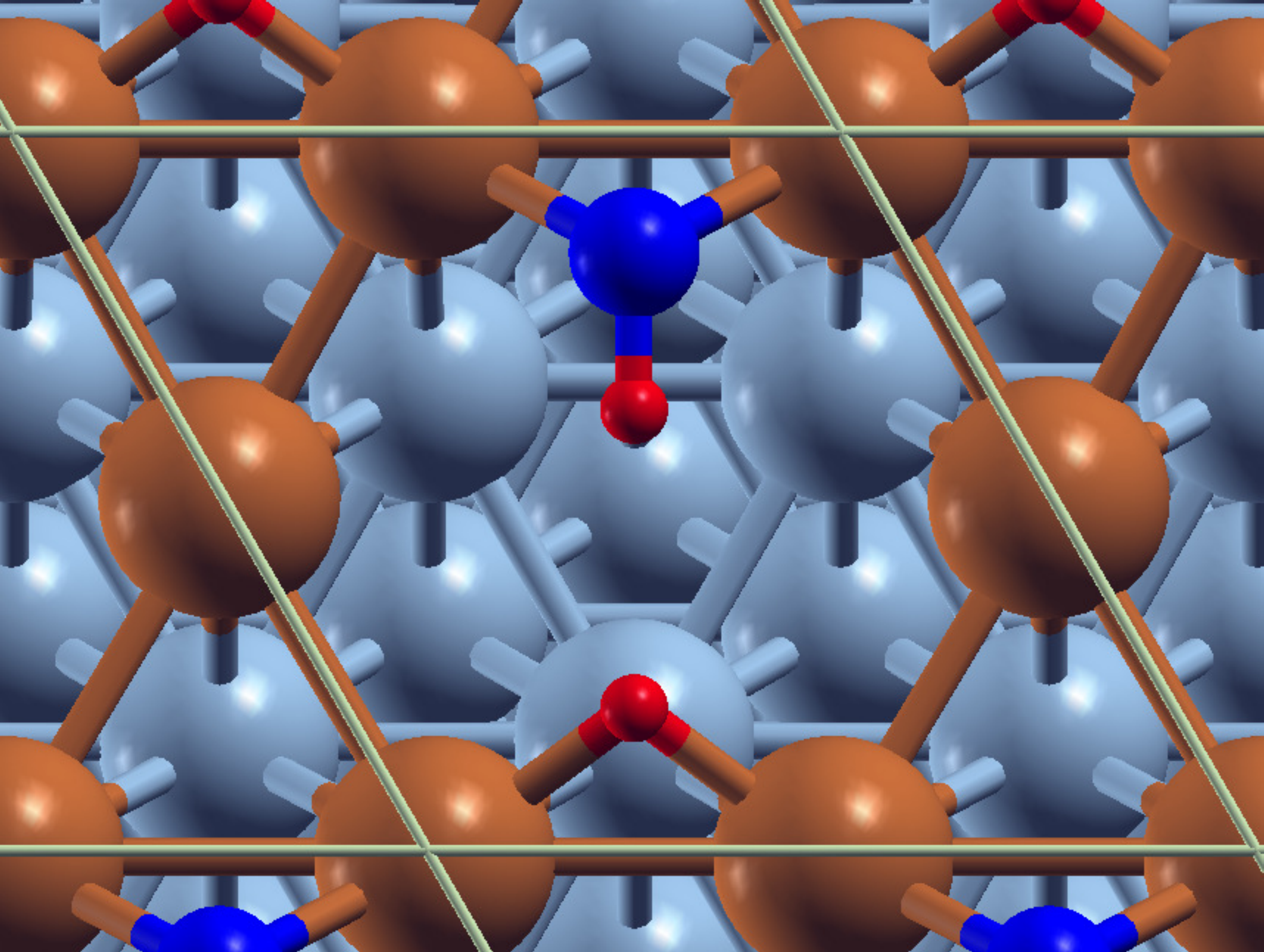}
  \label{fig:ohhlocoord}
  }
  \caption{(Color online) Representative configurations of H+OH+\slab~for three types of FSs.
  Uppermost layer of Cu atoms is
  represented by brown colored spheres, while all Cu atoms in layers below are
  colored grey, for visual aid.
  H atoms are represented as red spheres, and O atoms as blue spheres.
  (a) The most prevalent, most stable configuration where H  and  OH bind at the
  pristine Cu site, around the vacancy. (b) Configuration in which H  and  OH
  coordination is maximized by distorting upper Cu-layer. (c) Low coordinated OH
   and  H, with HO parallel to the Cu(111) plane. }
  \label{fig:FSconfigs}
\end{figure*}

In order to arrive at conclusions about dissociated water,
product atoms O, H, and H, were initially kept at different non-equivalent sites.
Their optimizations led to the most common dissociation product of H+OH.
Hence, OH was then
placed at the vacancy site, and H atom was placed at various remaining
positions, and optimized.
In another set of optimizations, OH was kept at the vacancy site, at a distance of $\sim$2 \AA~above 
the slab, and different H atom positions were tried. 
A total of 27 combinations were optimized, which
resulted into 16 distinct slab+products optimized geometries. These 16 distinct
geometries were either distinct in positions, and/or in energies. Their
dissociation energies spanned a range of about 0.6 eV.
A concise representation of prominent of these geometries, along with their energies is shown
in Fig. \ref{fig:vcnohhenergies}.
These 16 configurations can be divided into three distinct
classes based on their coordination with the surface.
First, and the most prevalent type of configurations of FS are the ones in which 
H+OH are along the vacancy boundary (See Fig. \ref{fig:ohhmostcommon}).
In this configuration, both, H and OH, sit at the pristine Cu site, maximizing their
coordination, with threefold equivalent bonds formed between H and three Cu atoms,
and HO and three Cu atoms, respectively. 
This is the most stable FS configuration among all other FS configurations, and
lies on the lower side of the graph in Fig. \ref{fig:vcnohhenergies}. 
In fact, the energy of this configuration is lower than the reactant,
slab+\water, by $\sim$ 0.12 eV.
The products avoid the vacancy in the most stable configuration, since their coordination 
cannot be increased at the vacancy site.
The next class of FS consists of configurations in which H and
OH maximize their coordination, by deforming the topmost layer
of Cu atoms, within plane. This class consists of more than one kind of
in-plane distortions of top Cu layers, in which the H and OH coordinate with either 2,
or 3 Cu-atoms of the topmost plane. A representative configuration is shown in
Fig. \ref{fig:ohhdistort}.
The last class of optimized structures consists of HO being parallel to the
Cu(111) plane, rather than being perpendicular to it, with lower (2-Cu) coordination.
A distinct stable FS configuration was also reached upon optimization, that
comprised of bound O atom, and adsorbed H$_2$ molecule.
It is the only one, out of the 27 optimized configurations, which dissociates water
in products other than H+OH.

Thus, symmetric arrangement of H+OH with 3 coordination is the most bound state
of the products, while in least bound state, OH gets oriented parallel to (111)
plane, with reduced coordination of 2. 

\subsection{NEB}
Various optimizations for the initial and the final states of dissociation reaction 
give a single initial state for the reaction, and 16 different possible final
states. NEB calculations are performed on 14 of these distinct final states. 
However, in some FS configurations, there were more than one possible ways in
which water molecule split to give that particular FS configuration.
Competing pathways of splitting \water, in such cases, did produce different
activation energies during CINEB runs.
We find that the activation energy for dissociation of water on Cu(111) slab with
vacancy, is most commonly confined to range 1.04 -- 1.18 eV.
The FS in which H  and  OH are low-coordinated with two Cu atoms,
possesses the least of activation energy (Top right part of Fig. \ref{fig:vcnohhenergies}), 
while the most stable of FS, in which H and  OH form
bonds with 3 Cu atoms along the periphery of vacancy, has the highest activation
energy (See Fig. \ref{fig:ohhmostcommon}).

Analysing the transition states (TSs) led us to the conclusion that there was a
common trend in all these different MEPs. The activation energy in all the MEPs
comprised of the step where water molecule on the slab gets closer to the slab,
than their equilibrium distance, and attached itself to one of the Cu atoms.
TS occurred in configuration in which actual splitting of this slab-bound water
molecule takes place. Thus, the initial climb of the activation energy was found to be the
same for all the MEPs.
Since IS for all these CINEB calculations is identical, and also the MEP
corresponding to activation energy climb, forward activation energies could be
expected to be identical in all these calculations. 
Indeed, the forward activation energies varied only in a small range of 0.14 eV.
Although the MEP of activation energy climb is similar for all the CINEBs, 
binding of water to the surface differs slightly with respect to the vacancy
site, and the Cu atom positions. This is what introduces the variation of 0.14
eV in forward activation energy in spite of same IS and similar forward MEP.

\begin{figure}[!h]
 \centering
 \includegraphics[scale=0.70] {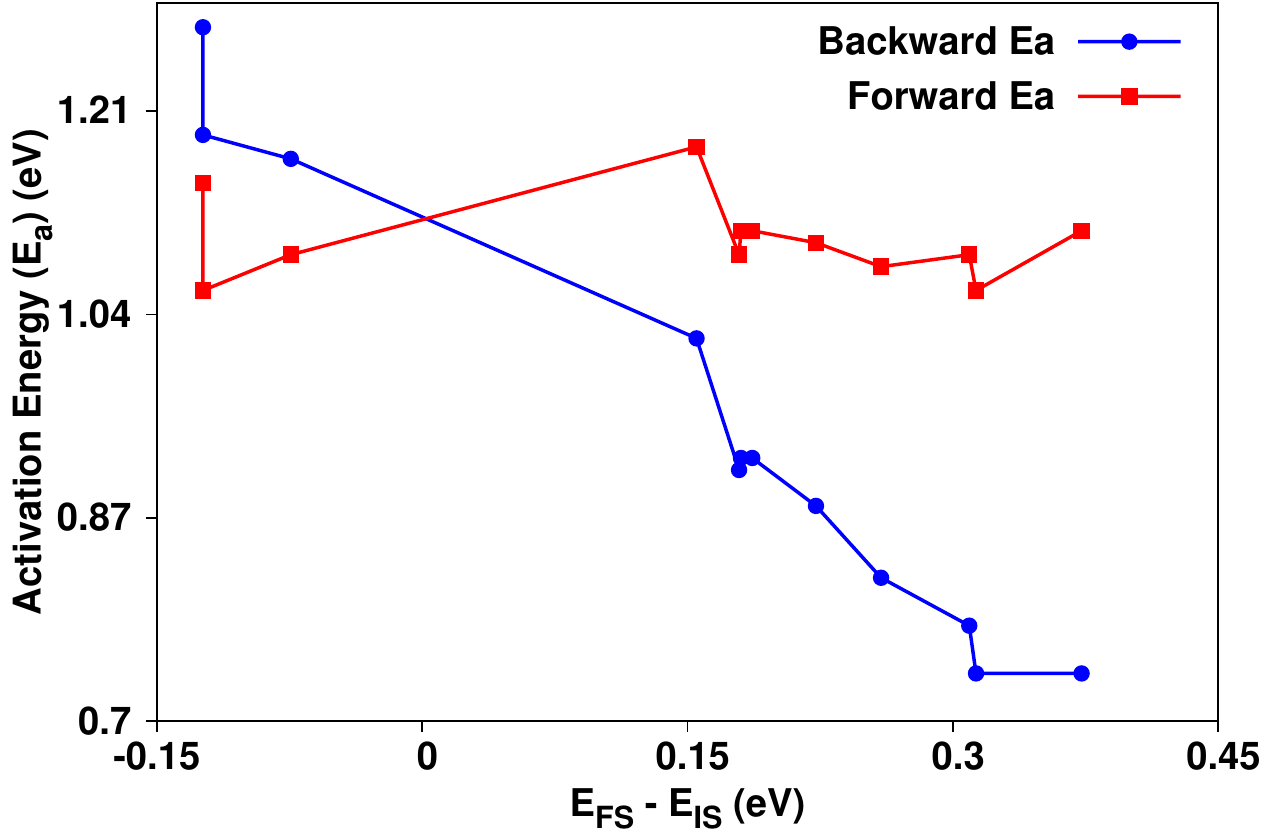}
 \label{fig:b}
  \caption{(Color online) Forward and backward activation energy as a function of energy of FS
  configuration with respect to the IS energy. While forward activation energy
  (square points in red color)
  is almost constant for different FSs, a systematic decrease in backward
  activation energies (filled circles in blue color) is observed with increasing energy of FS. Higher the
  energy of the FS, lower is the binding of product moieties with the
  Cu(111)+vacancy slab.}
  \label{fig:eavsene}
\end{figure}

Dissociation energies of the multiple FS configurations varied in the range of
0.6 eV. Since FS is not constant, unlike IS,
variation in the FS's energies, and configurations, is reflected in
an equally wide range in backward activation energy. Backward activation energy
exhibits a marked trend with respect to (wrt) the energy of the FS.
A graph of backward activation energies against the energy of the FS 
is shown in Fig. \ref{fig:eavsene}. 
Lower the energy of FS, harder it is to reform \water~from the dissociated H+OH
moieties. Hence, lower the energy of FS, greater will be the energy of
activation to form \water~from H+OH, and viceversa.
Lowering of coordination of the
dissociated H+OH products lowers their binding to the surface, also facilitating their easier
removal from the surface for reactions that have water dissociation as an
intermediate step.
\begin{table}[!h]
   \begin{tabular}{
    m{1.1cm} | m{0.42cm} m{0.65cm} |  m{0.42cm}    m{0.42cm}
    m{0.47cm}|m{0.42cm}  m{0.42cm} m{0.42cm}} \hline
              &                  &             & \multicolumn{6}{c}{Distance (\AA)} \\
 \cline{4-9}
 E$_{FS}$ (eV) & \multicolumn{2}{c|}{E$_a$(eV)} & \multicolumn{3}{c|}{HO-Cu } & \multicolumn{3}{c}{H-Cu} \\
 \cline{2-9}
  & E$_a^{Fwd}$  & E$_{a}^{Bkwd}$  & 1  &  2  &  3   & 1 & 2 & 3  \\
    \hline
              \multicolumn{9}{c}{Symmetric high coordinated FS}  \\
    \hline
    -0.1238      & 1.15      &  1.28         & 2.04  & 2.04  &  2.04  & 1.73  &  1.74  & 1.74 \\
    -0.1238      & 1.06      &  1.19         & 2.04  & 2.04  &  2.04  & 1.73  & 1.74   & 1.74 \\
    -0.0742      & 1.09      &  1.17         & 2.05  & 2.05  &  2.05  &  1.73  &  1.73  & 1.73 \\
    \hline
              \multicolumn{9}{c}{Surface distorting FS}  \\
    \hline
     0.1551      & 1.18      &  1.02         & 2.14  & 2.14  &  2.14  &  1.76  &  1.76  & 1.77 \\
     0.1791      & 1.09      & 0.91     & 2.14  & 2.15  &  2.16  &  1.71  &  1.79  & 1.82 \\
     0.1801      & 1.11      &  0.92         & 2.16  & 2.16  &  2.16  &  1.72  &  1.78  & 1.81 \\
     0.1865      & 1.11      &  0.92         & 2.16  & 2.17  &  2.17  &  1.74  &  1.76  & 1.78 \\
     0.2226      & 1.10      &  0.88         & 2.04  & 2.12  &  2.15  &  1.61  &  1.69  &  --  \\
     0.2593      & 1.08      &  0.82         & 2.00  & 2.00  &   --   &  1.64  &  1.64  &  --  \\
    \hline
              \multicolumn{9}{c}{Low coordinated products' FS}  \\
    \hline
     0.3093      & 1.09      &  0.78         & 1.97  & 1.98  &   --   &  1.65  &  1.65  &  --  \\
     0.3131      & 1.06 &  0.74    & 2.01  & 2.01  &   --   &  1.64  &  1.64  &  --  \\
     0.3727      & 1.11 &  0.74    & 1.98  & 1.99  &   --   &  1.64  &  1.65  &  --  \\
    \hline
   \end{tabular}
   \caption{Changes in bondlengths and activation energies (\ea) with decreasing
   binding between dissociated products and the Cu atoms of the Cu(111) slab + vacancy.}
   \label{table:withkpt}
\end{table}
The classification of FS configurations, discussed earlier in section
\ref{subsec:classifyFS}, also fits well with this variation.
We see that all FSs in which H  and  OH bind with two Cu atoms, 
possess the least (backward) activation
energy (0.33 -- 0.78 eV).  While, all FS in which the
products H  and  OH are coordinated with 3 Cu atoms via distortion of the uppermost Cu
surface, (0.82 -- 1.02 eV) lie in between. The highest backward activation energies are
exhibited by FS that possess the highest coordination, and high symmetry,
bonding with 3 Cu atoms along the boundary of the vacancy (1.17--1.28 ev).
Thus, low coordinated product states
are conducive to lowering of the backward activation energy, while more symmetric
arrangement of products corresponds to higher activation energy. 

\begin{figure}[!h]
  \centering
  \includegraphics[scale=0.70] {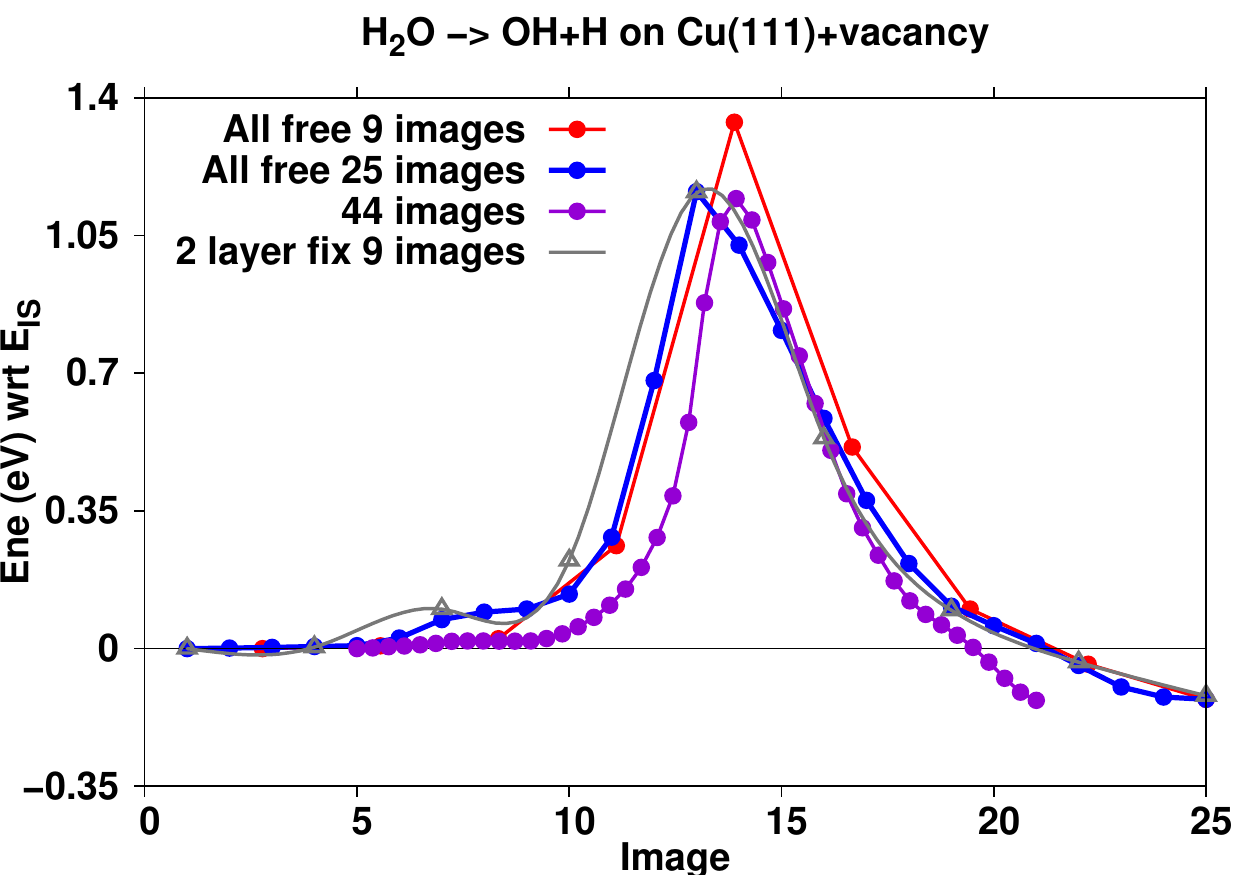}
  \caption{(Color online)
Increasing the intermediate images lowers \ea~ from 1.33 eV to 1.16 eV. 
Increasing intermediate images further did not changed the \ea~ from 1.16 eV .
}
  \label{fig:allmep.11.10-1}
\end{figure}

\begin{figure}[!h]
  \includegraphics[scale=0.70] {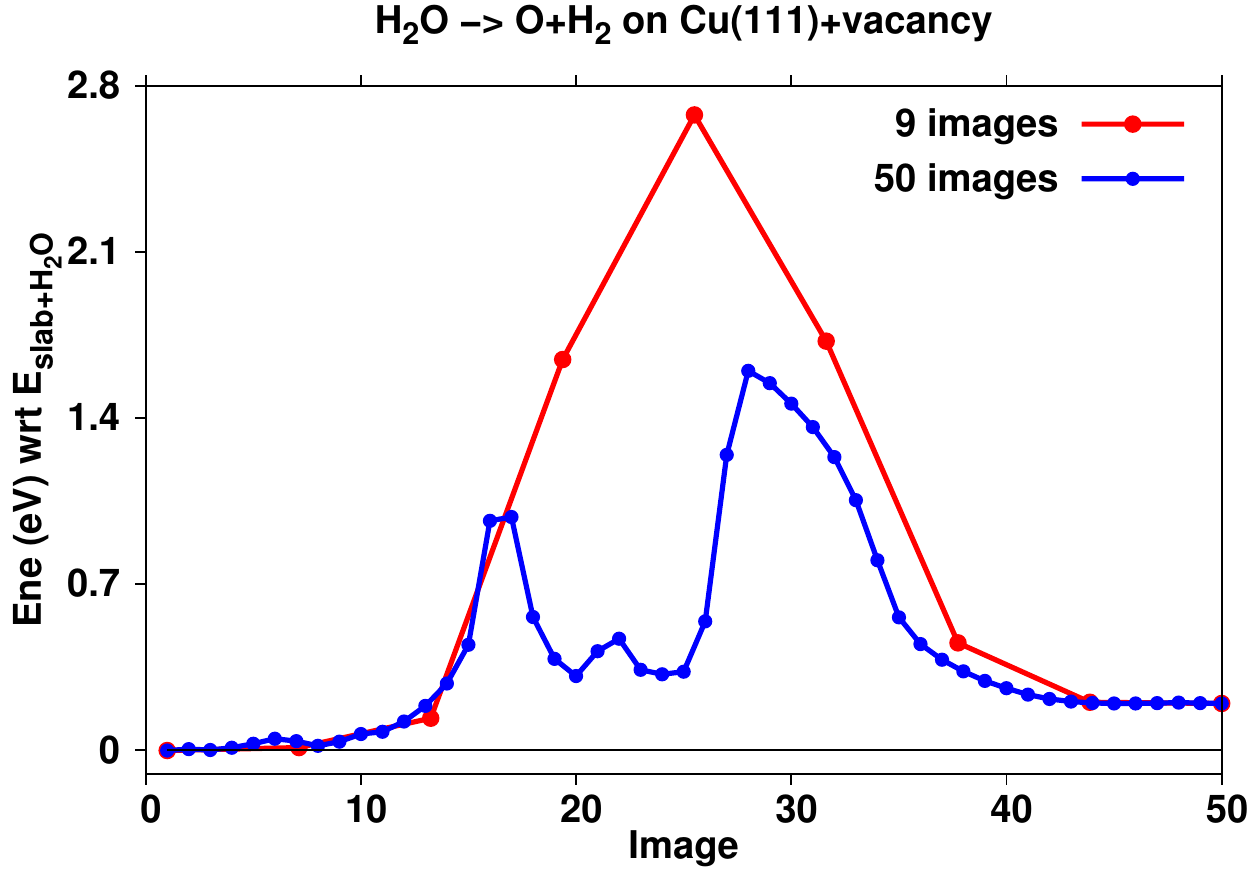}
  \caption{(Color online) MEP of reaction in which water dissociated initially into O+\ce{H2}.
  Increasing the number of images shows that the reaction is a two step process.
  First step in which \water~dissociates into H+OH
  (First peak 1.0 eV), and then OH further dissociates into O+H,
  (Second peak 1.61 eV), so that ultimately, O ends up bound to the surface, and
  free \ce{H2} is formed in the final state.}
\label{fig:allmep3.14}
\end{figure}

Further CINEB runs were performed for a few representative MEPs,
with larger number of images ($>$20) in order to support our conclusions.
MEP of FS geometry with symmetric 3 Cu coordinated OH and H was recalculated
using varying number of images (25, 44 images, and partially fixed slab
with 9 images).
Activation energy of this configuration changed from 1.33 eV with 9 images, to
1.16 eV, when 25, as well as 44 images were used.
Activation energy turned out to be 1.16 eV when 9 images were used with
partially fixed slab (See Fig. \ref{fig:allmep.11.10-1}).
Along with the symmetric FS's NEB,
a geometry in which OH  and  H exhibited low coordination (2) was also
repeated with increased number of NEB images.
It was chosen to represent the low coordination FS class.
A CINEB with 50 images was
performed as an extension to the already performed CINEB of 9 images. 
We found that its activation energy remained unchanged upon increasing the
number of images from 9 to 50. 
Interestingly, the reaction in which \water~dissociates into \ce{O + H2}, exhibited
a drastic change in activation energy from 2.67 eV to 1.61 eV, when number of images was increased
from 9 to 50. The 9-image MEP became a two step process when 50 images were used. 
The first step in the newly found MEP
involved dissociation of \water~into H+OH with energy of $\sim$ 1 eV, similar
to other \ce{H2O -> H + OH} MEPs. The later part of the 50-image MEP
involved breaking of OH bond while O remaining attached to the Cu atom, 
into H and  O, so that finally \ce{H2 + O} products are formed. Activation energy
of the second part of MEP was 1.61 eV (See Fig. \ref{fig:allmep3.14}).
\section{conclusions}
In this computational exercise we have demonstrated that 
introduction of vacancy defect in the Cu(111) pristine surface is one 
possible way to reduce the activation energy of water dissociation.
Activation energy of water dissociation can be lowered
by as much as 20\% with introduction of a vacancy on the otherwise pristine Cu(111) surface.
Weak interaction of water molecule with the copper surface is largely unaffected
by the introduction of vacancy, which reflects in `almost constant' forward activation
energy that differs only slightly for different MEPs. The backward activation energies show a
marked trend of decrease with lowering of coordination of the product moieties.
We have also shown that breaking the symmetry of pristine surface
by vacancy formation gives way to surface reconstruction,
which opens up a new possibility to lower the activation
energies of the reaction, that is otherwise absent in pristine surfaces.
Activation energy of the route of `surface
reconstruction' is intermediate in terms of binding energy, and activation
energy, between the low
coordinated low binding product states, and high symmetry higher coordinated
product states.
We find that activation energy is a result of the adsorbed water molecule approaching the
slab, closer than its equilibrium distance, and getting chemically bonded to the
surface. Transition state is the configuration in which water splits at this
attachment onto the slab.
CINEB calculations carried out with large number of images showed that
breaking of water into \ce{O + H2} is a two step process that involves
dissociation of water into \ce{H + OH} as an intermediate step. 

In view of the above results, we feel that the effect of density of vacancy upon activation
energy is a worthy investigation for future work.

\section{Acknowledgements}
The authors thank MSM (CSC-0129) for the financial support.
CSIR-4PI and CDAC are gratefully acknowledged for availing their computational
facility, used to carry out this work.
\bibliography{bibliography,vasp}

\begin{thebibliography}{10}

\bibitem{waterincatalysisbook}
Ib~Chorkendorff and Johannes~W Niemantsverdriet.
\newblock {\em Concepts of modern catalysis and kinetics}.
\newblock John Wiley \& Sons, 2006.

\bibitem{wgskineticsreview}
RJ~Smith, Muruganandam Loganathan, Murthy~Shekhar Shantha, et~al.
\newblock A review of the water gas shift reaction kinetics.
\newblock {\em International Journal of Chemical Reactor Engineering}, 8(1),
  2010.

\bibitem{wgsmicrokinetic}
CV~Ovesen, BS~Clausen, BS~Hammersh{\o}i, G~Steffensen, T~Askgaard,
  Ib~Chorkendorff, Jens~Kehlet N{\o}rskov, PB~Rasmussen, Per Stoltze, and
  P~Taylor.
\newblock A microkinetic analysis of the water--gas shift reaction under
  industrial conditions.
\newblock {\em Journal of Catalysis}, 158(1):170--180, 1996.

\bibitem{methanolfundamentals}
Alexander~Ya Rozovskii and Galina~I Lin.
\newblock Fundamentals of methanol synthesis and decomposition.
\newblock {\em Topics in Catalysis}, 22(3-4):137--150, 2003.

\bibitem{methanolsteamreformationreview}
Sandra S{\'a}, Hugo Silva, L{\'u}cia Brand{\~a}o, Jos{\'e}~M Sousa, and
  Ad{\'e}lio Mendes.
\newblock Catalysts for methanol steam reforming--{A} review.
\newblock {\em Applied Catalysis B: Environmental}, 99(1):43--57, 2010.

\bibitem{h2obioball}
Philip Ball.
\newblock Water as an active constituent in cell biology.
\newblock {\em Chemical Reviews}, 108(1):74--108, 2008.

\bibitem{bagchi2013waterbook}
Biman Bagchi.
\newblock {\em Water in Biological and Chemical Processes: From Structure and
  Dynamics to Function}.
\newblock Cambridge University Press, 2013.

\bibitem{h2obioproc}
Philippa~M Wiggins.
\newblock Role of water in some biological processes.
\newblock {\em Microbiological Reviews}, 54(4):432--449, 1990.

\bibitem{natureartificialphotosynthesis}
Artificial photosynthesis for solar water-splitting.
\newblock {\em {N}aure Photonics}, 6:511--518, 2012.

\bibitem{naturealtenergy}
M.~S. Dresselhaus and I.~L. Thomas.
\newblock Alternative energy technologies.
\newblock {\em Nature}, 414:332--337, 2001.

\bibitem{solarwatersplit}
Michael~G Walter, Emily~L Warren, James~R McKone, Shannon~W Boettcher, Qixi Mi,
  Elizabeth~A Santori, and Nathan~S Lewis.
\newblock Solar water splitting cells.
\newblock {\em Chemical Reviews}, 110(11):6446--6473, 2010.

\bibitem{inorganicwatersplit}
Frank~E Osterloh.
\newblock Inorganic nanostructures for photoelectrochemical and photocatalytic
  water splitting.
\newblock {\em Chemical Society Reviews}, 42(6):2294--2320, 2013.

\bibitem{naturephotocells}
Michael Gr\"atzeil.
\newblock Photoelectrochemical cells.
\newblock {\em Nature}, 414:338--344, 2001.

\bibitem{watermetalifacenature}
Javier Carrasco, Andrew Hodgson, and Angelos Michaelides.
\newblock A molecular perspective of water at metal interfaces.
\newblock {\em {N}ature Materials}, 11(8):667--674, 2012.

\bibitem{gokhalecu111wgs}
Amit~A Gokhale, James~A Dumesic, and Manos Mavrikakis.
\newblock On the mechanism of low-temperature water gas shift reaction on
  copper.
\newblock {\em Journal of the American Chemical Society}, 130(4):1402--1414,
  2008.

\bibitem{watersurfthiel}
Patricia~A Thiel and Theodore~E Madey.
\newblock The interaction of water with solid surfaces: Fundamental aspects.
\newblock {\em Surface Science Reports}, 7(6):211--385, 1987.

\bibitem{watersurfrevisited}
Michael~A Henderson.
\newblock The interaction of water with solid surfaces: fundamental aspects
  revisited.
\newblock {\em Surface Science Reports}, 46(1):1--308, 2002.

\bibitem{wateradswettingsurfrevue}
A~Hodgson and S~Haq.
\newblock Water adsorption and the wetting of metal surfaces.
\newblock {\em Surface Science Reports}, 64(9):381--451, 2009.

\bibitem{h2oonsi001}
Jun-Hyung Cho, Kwang~S Kim, Sung-Hoon Lee, and Myung-Ho Kang.
\newblock Dissociative adsorption of water on the {S}i(001) surface: {A}
  first-principles study.
\newblock {\em Physical Review B}, 61(7):4503, 2000.

\bibitem{nooxidationsteps}
CP~Vinod, JW~Niemantsverdriet, and BE~Nieuwenhuys.
\newblock Interaction of small molecules with {A}u (310): Decomposition of
  {NO}.
\newblock {\em Applied Catalysis A: General}, 291(1):93--97, 2005.

\bibitem{cooxidationaunp}
Ioannis~N Remediakis, Nuria Lopez, and Jens~K N{\o}rskov.
\newblock {CO} oxidation on gold nanoparticles: Theoretical studies.
\newblock {\em Applied Catalysis A: General}, 291(1):13--20, 2005.

\bibitem{langmuir}
Michael~A Henderson.
\newblock Structural sensitivity in the dissociation of water on {TiO$_2$}
  single-crystal surfaces.
\newblock {\em Langmuir}, 12(21):5093--5098, 1996.

\bibitem{o2dissociation}
Jos{\'e}~LC Faj{\'\i}n, M~Nat{\'a}lia~DS Cordeiro, and Jos{\'e}~RB Gomes.
\newblock Adsorption of atomic and molecular oxygen on the {A}u(321) surface:
  {DFT} study.
\newblock {\em The Journal of Physical Chemistry C}, 111(46):17311--17321,
  2007.

\bibitem{waterflat}
A~Michaelides, VA~Ranea, PL~De~Andres, and DA~King.
\newblock General model for water monomer adsorption on close-packed transition
  and noble metal surfaces.
\newblock {\em Physical Review Letters}, 90(21):216102, 2003.

\bibitem{h2ocu111phatak}
Abhijit~A Phatak, W~Nicholas Delgass, Fabio~H Ribeiro, and William~F Schneider.
\newblock Density functional theory comparison of water dissociation steps on
  {Cu, Au, Ni, Pd, and Pt}.
\newblock {\em The Journal of Physical Chemistry C}, 113(17):7269--7276, 2009.

\bibitem{h2ocu111disspnas}
Bin Jiang, Xuefeng Ren, Daiqian Xie, and Hua Guo.
\newblock Enhancing dissociative chemisorption of \ce{H2O} on \ce{Cu (111)} via
  vibrational excitation.
\newblock {\em Proceedings of the National Academy of Sciences},
  109(26):10224--10227, 2012.

\bibitem{wateroncustepsfajin}
Jos{\'e}~LC Faj{\'\i}n, M~Nat{\'a}lia~DS Cordeiro, Francesc Illas, and
  Jos{\'e}~RB Gomes.
\newblock Influence of step sites in the molecular mechanism of the water gas
  shift reaction catalyzed by copper.
\newblock {\em Journal of Catalysis}, 268(1):131--141, 2009.

\bibitem{paw1}
P.~E. Bl\"ochl.
\newblock Projector augmented-wave method.
\newblock {\em Physical Review B}, 50:17953--17979, Dec 1994.

\bibitem{paw2}
G.~Kresse and D.~Joubert.
\newblock From ultrasoft pseudopotentials to the projector augmented-wave
  method.
\newblock {\em Physical Review B}, 59:1758--1775, Jan 1999.

\bibitem{pbe1}
John~P. Perdew, Kieron Burke, and Matthias Ernzerhof.
\newblock Generalized gradient approximation made simple.
\newblock {\em Physical Review Letters}, 77:3865--3868, Oct 1996.

\bibitem{pbe2}
John~P. Perdew, Kieron Burke, and Matthias Ernzerhof.
\newblock Generalized gradient approximation made simple [{Phys. Rev. L}ett.
  77, 3865 (1996)].
\newblock {\em Physical Review Letters}, 78:1396--1396, Feb 1997.

\bibitem{quantumespresso}
Paolo Giannozzi, Stefano Baroni, Nicola Bonini, Matteo Calandra, Roberto Car,
  Carlo Cavazzoni, Davide Ceresoli, Guido~L Chiarotti, Matteo Cococcioni,
  Ismaila Dabo, et~al.
\newblock {QUANTUM ESPRESSO}: a modular and open-source software project for
  quantum simulations of materials.
\newblock {\em Journal of Physics: Condensed Matter}, 21(39):395502, 2009.

\bibitem{vdw1}
M.~Dion, H.~Rydberg, E.~Schr\"oder, D.~C. Langreth, and B.~I. Lundqvist.
\newblock Van der {W}aals density functional for general geometries.
\newblock {\em Physical Review Letters}, 92:246401, Jun 2004.

\bibitem{vdw2}
T.~Thonhauser, Valentino~R. Cooper, Shen Li, Aaron Puzder, Per Hyldgaard, and
  David~C. Langreth.
\newblock Van der {W}aals density functional: Self-consistent potential and the
  nature of the van der {W}aals bond.
\newblock {\em Physical Review B}, 92:125112, Sep 2007.

\bibitem{vdw3}
Guillermo Rom\'an-P\'erez and Jos\'e~M. Soler.
\newblock Efficient implementation of a van der {W}aals density functional:
  Application to double-wall carbon nanotubes.
\newblock {\em Physical Review Letters}, 103:096102, Aug 2009.

\bibitem{cualatwy}
Ralph Wyckoff.
\newblock {WG} crystal structures.
\newblock {\em New York, Interscience Publishers}, 1:7--83, 1963.

\bibitem{neb1}
Daniel Sheppard, Rye Terrell, and Graeme Henkelman.
\newblock Optimization methods for finding minimum energy paths.
\newblock {\em The Journal of Chemical Physics}, 128(13):134106, 2008.

\bibitem{neb2}
Graeme Henkelman and Hannes J{\'o}nsson.
\newblock Improved tangent estimate in the nudged elastic band method for
  finding minimum energy paths and saddle points.
\newblock {\em The Journal of Chemical Physics}, 113(22):9978--9985, 2000.

\bibitem{cineb}
Graeme Henkelman, Blas~P Uberuaga, and Hannes J{\'o}nsson.
\newblock A climbing image nudged elastic band method for finding saddle points
  and minimum energy paths.
\newblock {\em The Journal of Chemical Physics}, 113(22):9901--9904, 2000.

\end{thebibliography}
\bibliographystyle{unsrt}
\end{document}